\documentstyle[12pt]{article}
\newcommand{\li}{\limits}
\begin{document}

\begin{center}
{\bf Uniqueness of a Negative Mode About a Bounce Solution}

\vspace{0.8cm} {
 Michael Maziashvili }

\vspace{0.5cm}\baselineskip=14pt

\vspace{0.5cm} \baselineskip=14pt {\it Department of Theoretical
Physics, Tbilisi State University, 380028 Tbilisi, Georgia}
\end{center}
\vspace{0.5cm}
\begin{abstract}
We consider the uniqueness problem of a negative eigenvalue in the
spectrum of small fluctuations about a bounce solution in a
multidimensional case. Our approach is based on the concept of
conjugate points from Morse theory and is a natural generalization
of the nodal theorem approach usually used in one dimensional
case. We show that bounce solution has exactly one conjugate point
at $\tau=0$ with multiplicity one.
\end{abstract}
In the leading semiclassical approximation tunneling transitions
are associated with classical solutions of Euclidean equations of
motion. A special type of such solutions, time-reversal invariant
solution that approaches the local minimum $\vec{q}_f$ of a
potential $V(\vec{q})$ at infinity referred to as a bounce,
dominates the WKB transition rate from the bottom of potential
well. In the quantum version of the theory, the classically stable
equilibrium state $\vec{q}_f$ becomes unstable through barrier
penetration. It is a false vacuum state. Callan and Coleman
\cite{Cal-Col} approached the problem of false vacuum decay by
evaluating the Euclidean (imaginary time) functional integral at
the bounce solution in the semiclassical (small-$\hbar$) limit.
They found that in this way the negative eigenvalue of the second
variational derivative of action at bounce makes an imaginary
energy shift and may be interpreted as a decay rate. Since, this
interpretation relies heavily upon the existence of an unique
negative mode, it is crucial to show that the second variational
derivative of action at the bounce has one and only one negative
eigenvalue. The negative mode problem was already considered by
Coleman \cite{Col-88}. We reconsider this problem by using the
concept of conjugate points. Discussion presented here has a
potential application in quantum field theory viewing a field
$\phi (\vec{x})$ as a collection of mechanical variables
$q^i~(i=1,\ldots ,N)$ for $N$ degrees of freedom, in the limit $N$
becomes continuously infinite: $\vec{q}\rightarrow
\phi(\vec{x})~,i\rightarrow \vec{x}.$ However, to take this limit
on a quite sound mathematical ground requires the use of a
rigorous functional analysis. Throughout this paper we will
restrict ourselves to the consideration of a multidimensional
case. We shall assume that the initial point of tunneling
$\vec{q}_{in}$ is taken arbitrarily. In this case tunnelling is
described by the solution of the imaginary-time equations of
motion which begin at some position $\vec{q}_{es}\neq\vec{q}_{in}$
at rest and come to rest at time $T/2$ at $\vec{q}_{in}.$ For the
sake of convenience we take $V(\vec{q}_{in})=0.$ Thus, the
solution we are interested in is defined from this zero-energy
solution by the time reflection,
$\vec{q}_b(\tau)=\vec{q}_b(-\tau).$ (The suffix b denotes bounce
like solution, in a particular case $T=\infty$ we arrive at the
bounce.) By its definition, this is an even, zero-energy
stationary point of action,
\begin{equation}\label{symlimits}S_E[\vec{q}]=\int\li_{-T/2}\li^{T/2}d\tau
\left(\frac{m_{ik}(\vec{q})\dot{q}^i\dot{q}^k}{2}+V(\vec{q})\right),
\end{equation} with the boundary conditions $\vec{q}_b(\pm
T/2)=\vec{q}_{in}$. Where $m_{ik}(\vec{q})$ is some
positive-definite symmetric matrix and the summation convention
over repeated indices is used. The matrix $m_{ik}(\vec{q})$
defines the metric in a configuration space,
$(d\vec{q})^2=m_{ik}dq^idq^k=dq_kdq^k.$ The corresponding
imaginary-time equations of motion take the form
\begin{equation}\label{eqm} \frac{\delta^2 q^i}{\delta\tau
^2}-\frac{\partial V}{\partial q_i}=0.
\end{equation} Where, $\frac{\delta}{\delta\sigma}=\dot{q}^i\nabla_i=\dot{q}_i\nabla^i$ is a covariant
derivative along the vector $\dot{q}^i$ and $\nabla_i$ is the
covariant derivative with respect to $q^i$ compatible with metric
$m_{ik}.$ According to the formalism developed by Banks, Bender
and Wu \cite{Ban-Ben-Wu}, in the leading semiclassical
approximation the tunnelling probability is dominated by the
solution that minimizes the Jacobi type action. Now, in order to
define the corresponding Jacobi type action, we introduce a
parameter $\sigma$ along the path, $\vec{q}(\sigma),$ that
increases monotonically from $-T/2$ at initial point
$\vec{q}_{in}$ to $0$ at the escape one $\vec{q}_{es}$. Denoting
$\dot{\vec{q}}\equiv\frac{d\vec{q}(\sigma )}{d\sigma}$, the action
is given by the functional
\begin{equation}J_E[\vec{q}]=\int\li_{-T/2}\li^{0}d\sigma
\sqrt{2V(\vec{q})m_{ik}(\vec{q})\dot{q}^i\dot{q}^k},
\label{act}\end{equation} over trajectories $\vec{q}(\sigma)$
connecting two boundary points on different sides of the barrier,
$\vec{q}(-T/2)=\vec{q}_{in}$, and ,$\vec{q}(0)=\vec{q}_{es}.$ In
general the barrier penetration path has to be at least a local
minimum of this action. That is crucial for our further
discussion. An extremum of this action gives a classical path in
the configuration space of the system, but says nothing about its
motion in imaginary time. To determine the system's evolution in
imaginary time requires the use of a supplementary condition,
\begin{equation}\label{sc}
\frac{m_{ik}(\vec{q})}{2}\frac{dq^i}{d\tau}\frac{dq^k}{d\tau}-V(\vec{q})=0.
\end{equation}
The variational principles of mechanics are widely discussed in
Ref.\cite{Lan}. This extra relation is just the Euclidean energy
condition and once the configuration path $\vec{q}(\sigma )$ is
known, it can be integrated to get the imaginary time
parameterization $\sigma (\tau)$. The equation of motion following
from Jacobi type action (\ref{act}) and the supplementary
condition (\ref{sc}) is equivalent to imaginary time
equation (\ref{eqm}) with the Euclidean energy, a first integral of
Euclidean equation (\ref{eqm}), fixed to the value zero. This is
shown explicitly by varying (\ref{act}), which yields the equation
of motion
\begin{equation}\label{op}
\left(\frac{\delta^2{q^i}}{\delta\sigma^2}-\frac{\dot{\vec{q}}\,^2}{2V(\vec{q})}\,\nabla^iV\right)\Pi^k_i=0,
\end{equation}
where $\Pi^k_i=\delta^k_i-\dot{q}^k\dot{q}_i/\dot{\vec{q}}\,^2$ is
the projection operator onto the subspace of configuration space
that is orthogonal to the configuration space-path. If by using
Eq.(\ref{sc}) we parameterize a configuration space-path
$\vec{q}(\sigma )$ with parameter $\tau$ Eq.(\ref{op}) just
becomes
\begin{equation}\label{top}
\left(\frac{\delta^2{q^i}}{\delta\tau^2}-\nabla^iV\right)\Pi^k_i=0.
\end{equation}
Therefore, the equation of motion (\ref{op}) obtained from the
Jacobi type action (\ref{act}), supplemented by (\ref{sc}), is
equivalent to imaginary time equation of motion (\ref{eqm}) with
fixed zero Euclidean energy. The parallel projection of
Eq.(\ref{eqm}) to the configuration path follows from
Eq.(\ref{sc}), by differentiating with respect to $\tau$
\begin{equation}\label{pp}
\left(\frac{\delta^2{q^i}}{\delta\tau^2}-\nabla^iV\right)\frac{d
q_i }{d\tau}=0.
\end{equation}
Note that if we multiple Eq.(\ref{pp}) by the tangential vector
$\left.\frac{dq^k}{d\tau}\right/m_{nl}(\vec{q})\frac{dq^n}{d\tau}\frac{dq^l}{d\tau}$
and add to the Eq.(\ref{top}) we get the Eq.(\ref{eqm}). As an
essential point for our discussion we want to emphasize that the
Jacobi type action (\ref{act}) is invariant under the
reparametrizations of the configuration space-path that preserve
the end point values of the parameter. That is, (\ref{act}) is
invariant under the replacements $\sigma \rightarrow f(\sigma)$
and $q^i(\sigma)\rightarrow\bar{q}^i(f(\sigma))$ with
$f(-T/2)=-T/2$ and $f(0)=0$. Their infinitesimal form is $\sigma
\rightarrow\sigma+\epsilon(\sigma)$ and $q^i\rightarrow
q^i+\epsilon\dot{q}^i$ where $\epsilon(-T/2)=\epsilon(0)=0$. Due
to the Noether's second theorem, there is a corresponding gauge
identity of the form Eq.(\ref{pp}). Now it is obvious that the
proper fluctuations for Jacobi type action are transverse ones,
while the longitudinal ones reproduce a gauge transformation. Now
let us recall some definitions and statement from Morse theory
about the conjugate points. The Morse index for a given trajectory
is defined as a number of negative eigenvalues of a second
variation of action evaluated at this trajectory with the zero
boundary conditions at the endpoints. On the other hand due to
Morse theory the number of negative eigenvalues may be evaluated
by counting the conjugate points with their multiplicities
\cite{Morse}. The point $c$ is conjugate to $a$ with multiplicity
$m$ for the differential operator $A$ if the two-point boundary
value problem $Au=0,~u(a)=u(c)=0$ has the nontrivial $m$ linearly
independent solutions. The second variation of the action
Eq.(\ref{act}) in imaginary time parameterization gives the
following two-point boundary value problem,
\begin{equation}\label{sv} \left(-\frac{\delta^2\phi^i}{\delta\tau^2}-R^i_{\,jkl}\dot{q}_b^j\phi^k\dot{q}_b^l
+(\nabla^i\nabla_j
V)\phi^j\right)\Pi^k_i(b)=0,~\vec{\phi}(-T/2)=\vec{\phi}(0)={\bf
0}.
\end{equation}
Where $\Pi^k_i(b)$ denotes orthogonal projection onto the bounce
like solution. We are tacitly assuming here and below that the
terms $R^i_{\,jkl}$(the Riemann curvature tensor of a metric
$m_{ik}$), $\nabla^i\nabla_jV$ are evaluated at $\vec{q}_b$ and
$\frac{\delta}{\delta\tau}=\frac{dq^i_b}{d\tau}\nabla_i.$ The
barrier penetration path is at least a local minimum of action
Eq.(\ref{act}), i.e. all small transverse (proper) fluctuations
about this path increase this action. It means that ,
$\vec{\phi}$, must not contain a transverse part. On the other
hand, since the barrier penetration path is at least a local
minimum of action (\ref{act}) the second variation at this path
must be positive semidefinite operator, in such a case due to
Morse theory the $\vec{\phi}$ must be free of conjugate points.
Consider the following two-point boundary value problem,
\begin{equation}-\frac{\delta^2\psi^i}{\delta\tau^2}-R^i_{\,jkl}\dot{q}_b^j\psi^k\dot{q}_b^l
+(\nabla^i\nabla_j V)\psi^j=0,~\vec{\psi}(-T/2)=\vec{\psi}(0)={\bf
0},\label{esv}\end{equation} one concludes that the solutions
$\vec{\psi}$ automatically satisfy the two-point boundary value
problem (\ref{sv}). It means that $\vec{\psi}$ also must not
contain a transverse part and the conjugate points. Take a general
longitudinal ansatz satisfying the corresponding boundary
conditions $\vec{\psi}=\lambda(\tau)\frac{d\vec{q}_b}{d\tau},$
from Eq.(\ref{esv}) we get the following equation
$\frac{d}{d\tau}\left[\frac{d\lambda
(\tau)}{d\tau}\left(\frac{d\vec{q}_b}{d\tau}\right)^2\right]=0.$
It gives $\lambda=const.$ and correspondingly
$\vec{\psi}\sim\frac{d\vec{q}_b}{d\tau}.$ Combining these, we
obtain the following proposition.
\\{\bf Proposition 1.} All solutions of the two-point boundary
value problem (\ref{esv}) are linearly dependent and free of
conjugate points.\\ Now we turn our attention to a consideration
of the second variation of Eq.(\ref{symlimits}) evaluated at the
bounce like solution,
\begin{equation}\label{ze}
-\frac{\delta^2\chi^i}{\delta\tau^2}-R^i_{\,jkl}\dot{q}_b^j\chi^k\dot{q}_b^l
+(\nabla^i\nabla_j V)\chi^j=0,~\vec{\chi}(\pm T/2)={\bf 0}.
\end{equation} Since the solution $\vec{q}_b(\tau)$ is an even function
of $\tau,$ the second variational derivative in Eq.(\ref{ze})
commutes with the operator $T,~T:\vec{\chi}(\tau)=\vec{\chi}
(-\tau).$ Therefore, the function $\chi$ is to be either even or
odd with respect to $\tau$. Due to this symmetry the presence of
the conjugate point $\tau_{conj}
>0$ implies the presence of the second conjugate one $-\tau_{conj},$ but due to Proposition 1.
this is not the case. Nevertheless, the solution
$\vec{\chi}\sim\frac{d\vec{q}_b}{d\tau}$ has a conjugate point
$\tau=0.$ The multiplicity  of this conjugate point equals one.
Namely, from Proposition 1. we know that all solutions of
two-point boundary value problem Eq.(\ref{esv}) are linearly
dependent. Thus, we arrive at the following result.\\{\bf
Proposition 2.} The two-point boundary value problem Eq.(\ref{ze})
has exactly one conjugate point $\tau=0$ with multiplicity
one.\\Due to morse theory it means the presence of unique negative
mode in the spectrum of small fluctuations around the bounce
solution.

\section*{Acknowledgments}
I am very grateful to my advisor in Tbilisi State University,
Professor A.\,Khelashvili, for helpful conversations and to
Professor G.\,Lavrelashvili for useful discussions.

\end{document}